\documentclass[twoside,fleqn]{article}
\usepackage{espcrc2}
\usepackage{amssymb}
\usepackage[dvips]{epsfig}
\newcommand{\be}{\begin{equation}}
\newcommand{\ee}{\end{equation}}
\newcommand{\bea}{\begin{eqnarray}}
\newcommand{\eea}{\end{eqnarray}}
\newcommand{\bean}{\begin{eqnarray*}}
\newcommand{\eean}{\end{eqnarray*}}

\newcommand{\alphs}{\alpha_{\rm s}}
\newcommand{\alphSF}{\alpha_{\rm SF}}
\newcommand{\gbar}{\overline{g}}

\newcommand{\gbsq}{\overline{g}^2}
\newcommand{\gbsqi}{\overline{g}^{\,-2}}
\newcommand{\gbsqSF}{\overline{g}^2_{{\rm SF}}}

\newcommand{\lmax}{L_{\rm max}}

\title{
{
\vspace{-4.25cm} \normalsize \hfill
\parbox{21.5mm}{MS-TP-01-8\\DESY 01-151\\HU-EP-01/38}
}\\[30mm]
The Schr\"odinger functional coupling in quenched QCD at low energies%
\thanks{Based on a poster presented by J.H. at the conference
LATTICE '01, August 19 -- 24, 2001, in Berlin, Germany.}%
\thanks{Supported by the EU under HPRN-CT-2000-00145.}
}
\author{
Jochen Heitger\address{
WWU M\"unster, Institut f\"ur Theoretische Physik,
Wilhelm-Klemm-Str.~9, D-48149 M\"unster, Germany},
Hubert Simma\address{
Deutsches Elektronen-Synchrotron DESY Zeuthen, Platanenallee~6,
D-15738 Zeuthen, Germany},
Rainer Sommer$^{\mbox{\scriptsize b}}$ and
Ulli Wolff\,\address{
Humboldt Universit\"at Berlin, Institut f\"ur Physik,
Invalidenstr.~110, D-10099 Berlin, Germany}
(ALPHA Collaboration)
}
\begin{document}
%
%%%%%%%%%%%%%%%%%%%%%%%%%%%%%%%%%%%%%%%%%%%%%%%%%%%%%%%
% slight change in table style (J.H.,1996) %%%%%%%%%%%%
\makeatletter
\long\def\@maketablecaption#1#2{\vskip 10mm #1. #2\par}
\makeatother
%%%%%%%%%%%%%%%%%%%%%%%%%%%%%%%%%%%%%%%%%%%%%%%%%%%%%%%
% slight change in figure style (J.H.,2001) %%%%%%%%%%%
\makeatletter
\def\fnum@figure{Fig.~\thefigure}
\makeatother
%%%%%%%%%%%%%%%%%%%%%%%%%%%%%%%%%%%%%%%%%%%%%%%%%%%%%%%
%
\begin{abstract}
Existing non-perturbative computations of the running coupling of 
quenched QCD in the Schr\"odinger functional scheme are extended to 
scales $\mu$ lying much deeper in the low-energy regime.
We are able to reach $\mu^{-1}\approx 0.9\,{\rm fm}$, where a 
significant deviation from its perturbative evolution is observed.
\end{abstract}
\maketitle
%
%%%%%%%%%%%%%%%%%%%%%%%%%%%%%%%%%%%%%%%%%%%%%%%%%%%%%%%%%%%%%%%%%%%%%%%%%
%
\section{MOTIVATION}
The Schr\"odinger functional (SF) of QCD provides a suitable tool to 
compute the running of the strong coupling 
$\alphs$ \cite{gbar_SU3,gbar_Nf2} and to solve other renormalization 
problems \cite{ZP,NPren_lat00} non-perturbatively by means of numerical 
simulations.
The basic idea to cover the scales involved -- often differing by many 
orders of magnitude -- is a recursive finite-size scaling 
technique \cite{RFSST} to relate the low-energy sector with the scaling 
regime at high energies, where the evolution of $\alphs$ follows the 
perturbative renormalization group.

A particular property of the coupling in the SF scheme, $\alphSF$, which 
however should not be regarded as universal for QCD couplings, is that 
its non-perturbative running is quite accurately described by 
perturbation theory (PT) down to surprisingly low energies.
Therefore, it appears interesting to investigate $\alphSF(\mu)$ in 
quenched QCD at much lower scales $\mu=1/L$ ($L$ the linear size of the 
finite system) than considered so far \cite{gbar_SU3,ZP} in order to 
see a deviation from PT.

As an aside we would like to remark that during the year 2000 the 
present study at the same time was designed to test the APEmille 
computers installed at DESY Zeuthen \cite{APE_lat01} and to gain 
experiences with code adaptation and optimization for this architecture 
in the context of a first, clear-cut physics project.
\section{SF SETUP}
The (lattice regularized) SF is given in terms of the effective action
(free energy) $\Gamma$ of QCD satisfying Dirichlet boundary conditions 
in Euclidean time \cite{SF}.
A renormalized coupling, $\gbsqSF \equiv \gbsq$, is introduced as the 
response to an infinitesimal variation of a specific 1--parameter 
($\eta$) family of prescribed constant abelian boundary fields.
Taking into account the perturbative series
$\Gamma=\Gamma_0\,g_0^{-2}+\Gamma_1+\Gamma_2\,g_0^2+\cdots$, we define 
$\gbsq$ by
\be
\frac{\partial\Gamma}{\partial\eta}\,\bigg|_{\eta=0}=
\frac{\partial\Gamma_0}{\partial\eta}\,\bigg|_{\eta=0}\;
\frac{1}{\gbsq(L)}\,.
\ee

$\gbar$ genuinely depends only on one renormalization scale, $L=1/\mu$, 
and the quantity that has been devised to map out the scale evolution 
of $\gbar$ is the step scaling function (SSF) $\sigma(s,u)$:
\be
\sigma(s,u)=\gbsq(sL)\,\big|_{\gbsq(L)=u}\,.
\label{def_SSF}
\ee
It represents an integrated beta function for finite scale 
transformations with rescaling factor $s$ \cite{RFSST}.
In principle, if one has control over the exact SSF, one can trace the 
non-perturbative evolution of the coupling in discrete steps
$\gbsq(L) \to \gbsq(sL) \to \gbsq(s^2L) \to \cdots\,\,$.
For one rescaling step, $\sigma$ is computed as the continuum limit
$a\to 0$ of the SSF at finite resolution, $\Sigma(s,u,a/L)$.
So far the method of construction for each $u$--value has been to 
choose $L/a$, determine $\beta$ to match $\gbsq=u$ on the 
corresponding lattice and to then simulate on an $sL/a$--lattice
with the same $\beta$ and read off $\Sigma=\gbsq(sL)$.
For further details we refer to \cite{gbar_SU3,ZP}.
%
%%% Beginn Figur %%%
\begin{figure}[t]
\begin{center}
\epsfig{file=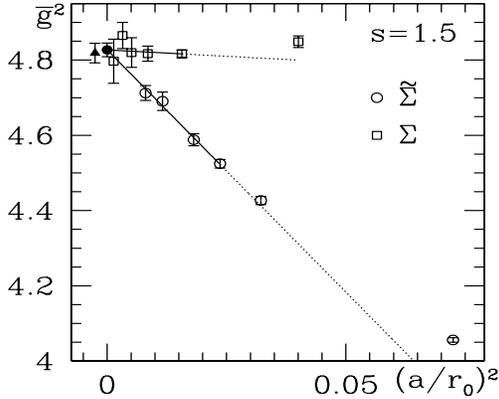,width=7.5cm,height=6.5cm}
\vspace{-2.0cm}
\caption[t_param]{\label{Sig348Fig} \sl
Joint continuum extrapolation of $\Sigma$ and $\tilde{\Sigma}$, 
and result of extrapolating $\tilde{\Sigma}$ alone (triangle).}
\end{center}
\vspace{-1.0cm}
\end{figure}
%%% Ende Figur %%%
%
\section{SIMULATIONS AND RESULTS}
\subsection{New simulation strategy}
\begin{figure}[t]
\begin{center}
\epsfig{file=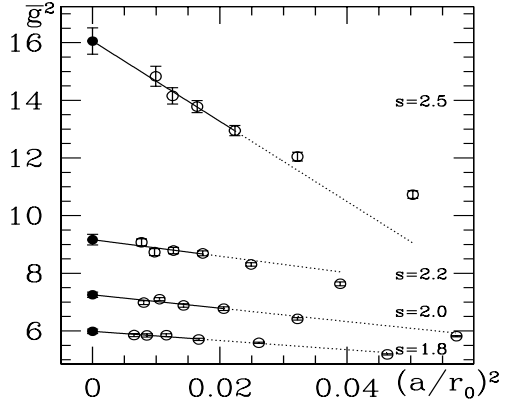,width=7.5cm,height=6.5cm}
\vspace{-2.0cm}
\caption[t_param]{\label{SigFig} \sl
Extrapolations of $\tilde{\Sigma}(s,3.48,a/L)$, discarding the two 
coarsest lattices.}
\end{center}
\vspace{-1.0cm}
\end{figure}

In our previous computations $\lmax$ defined by $\gbsq(\lmax)=3.48$ has 
been both an intermediate reference scale and the low energy bound of 
our scaling investigations.
In ref.~\cite{r0_ALPHA} this scale was connected with the scale
$r_0 \approx 0.5\,{\rm fm}$ from the static potential yielding
$\lmax/r_0=0.718(16)$ in the continuum limit.
In the same reference a relation connecting $\beta$ with $r_0/a$ was 
given for the range $5.7 \le\beta\le 6.57$.
These combined informations enable us to directly determine 
$\beta$--values corresponding to $L=s\lmax$ for a series of chosen 
resolutions $L/a$ for continuum extrapolation.
We just use the cited relation for $r_0/a=(r_0/\lmax)(L/a)(1/s)$ in 
the allowed range.
In this way we construct an alternative lattice estimator
$\tilde{\Sigma}(s,3.48,a/L)$ with the same continuum limit 
$\sigma(s,3.48)$ but cutoff effects different from $\Sigma$.
In fig.~\ref{Sig348Fig} we demonstrate consistency between the two 
approaches for $s=1.5$. 
In fig.~\ref{SigFig} continuum extrapolations of $\tilde{\Sigma}$  
are shown for other $s$--values, which now take us down to $2.5 \lmax$ 
in energy.

For the O($a$) boundary improvement term of the SF we use the 2--loop 
approximation \cite{ct_2loop}.
For our Monte Carlo simulations we employed a similar
`hybrid-overrelaxation' algorithm as in \cite{gbar_SU3}.
Each iteration consists of 1 heatbath update and $N_{\rm OR}$ 
subsequent overrelaxation sweeps (typically $N_{\rm OR}=5,7$).
As detailed there, non-gaussian tails and long autocorrelations in the
coupling at low energy are overcome by a modified sampling procedure
that enhances the tail contributions and is compensated by 
reweighting.
%
%%% Beginn Figur %%%
%%% Ende Figur %%%
%
\subsection{Results}
The SSF $\tilde{\Sigma}$--values in fig.~\ref{SigFig} approach the 
continuum at rate compatible with $a^2$.
This shows that lattice artifacts linear in $a$, which in the SF a 
priori exist, are invisible to our accuracy with 2--loop boundary 
improvement and allowed us to fit by the form 
$\tilde{\Sigma}=\sigma+\rho_2\,(a^2/r_0^2)$\,.
As a safeguard against underestimating the uncertainty in $\sigma$ 
we omitted the two coarsest lattices.
This is in contrast with the extrapolation of $\lmax/r_0$, where a 
small linear component was found \cite{r0_contd}.

%
%%% Beginn Figur %%%
\begin{figure}[htb]
\vspace{-1.125cm}
\begin{center}
\epsfig{file=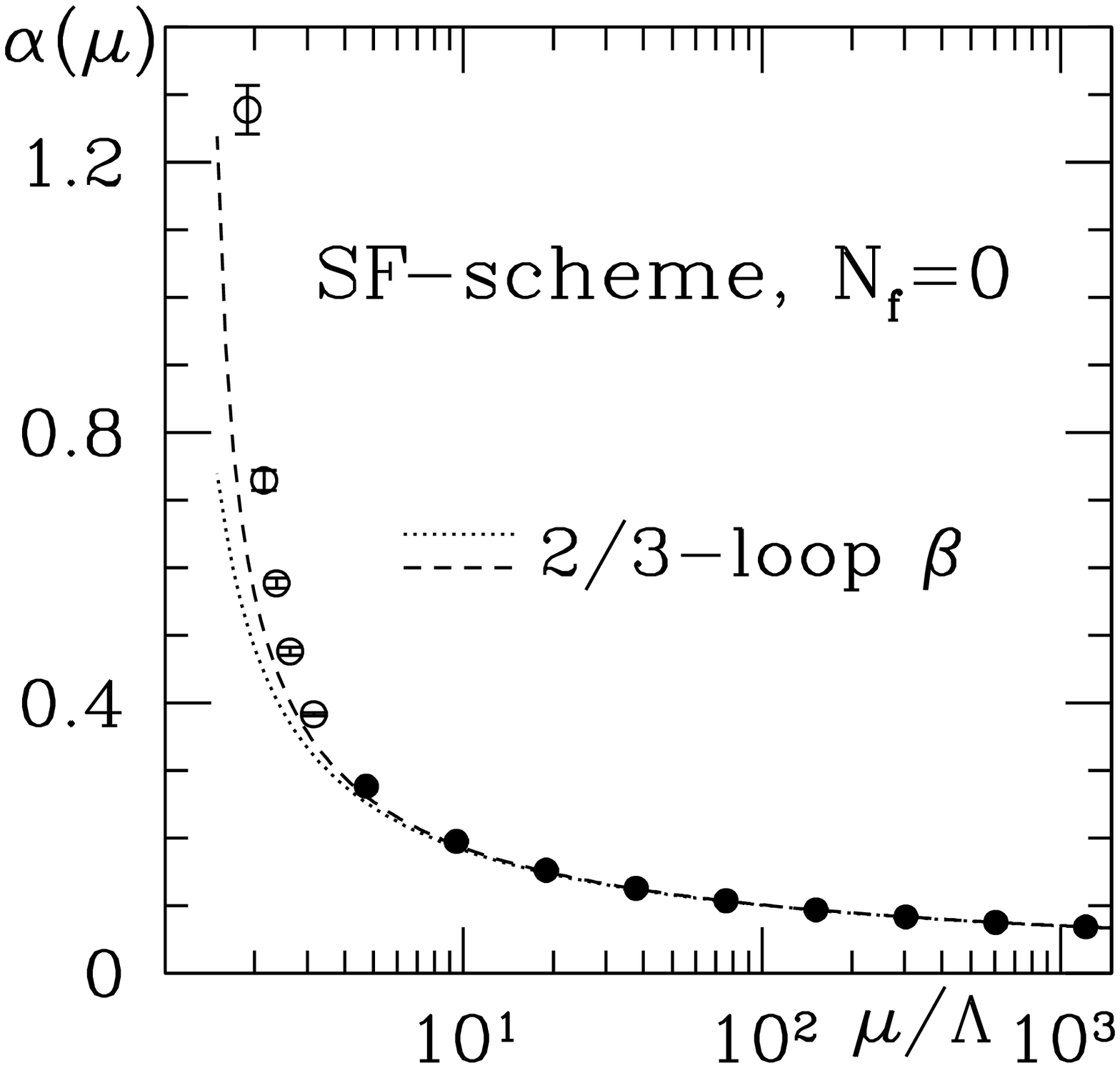,width=7.5cm,height=6.5cm}
\vspace{-2.0cm}
\caption[t_param]{\label{AlphaFig} \sl
Running of $\alpha\equiv\alphSF=\gbsq/4\pi$.}
\end{center}
\vspace{-1.5cm}
\end{figure}
%%% Ende Figur %%%
%
%%% Beginn Figur %%%
\begin{figure}[htb]
\begin{center}
\epsfig{file=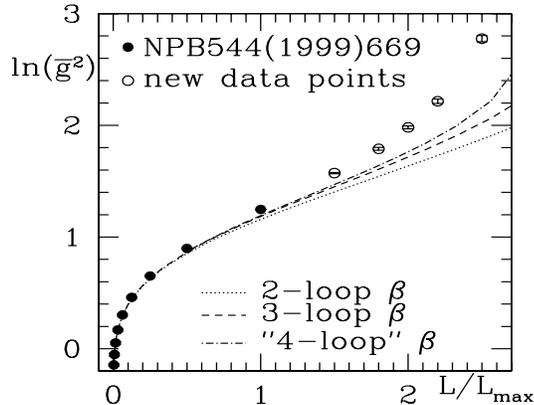,width=7.5cm,height=6.5cm}
\vspace{-2.125cm}
\caption[t_param]{\label{GbarFig} \sl
$L$--dependence of $\gbsq$ versus PT.
The artificial "4--loop" $\beta$--function uses $b_3=1/(4\pi)^4$.}
\end{center}
\vspace{-1.0cm}
\end{figure}
%%% Ende Figur %%%
%
Figs.~\ref{AlphaFig} and \ref{GbarFig} display the scale evolution of 
the SF coupling over the whole energy range that is available from 
refs.~\cite{gbar_SU3,ZP} together with this work and confront it with 
results from integrations with the perturbative $\beta$--function.
With decreasing energy a substantial deviation from PT now becomes 
clearly visible.
The steep growth of $\ln\{\gbsq(L)\}$ with $L$ reveals the 
non-perturbative behaviour to set in for $L>0.7\,r_0$.
\subsection{Discussion of the large--$L$ asymptotics}
%
%%% Beginn Figur %%%
\begin{figure}[t]
\vspace{-0.5cm}
\begin{center}
\epsfig{file=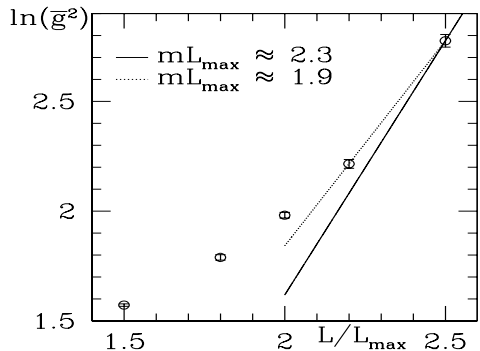,width=7.0cm,height=5.59cm}
\vspace{-1.625cm}
\caption[t_param]{\label{StrFig} \sl
Estimates of $m$ via leading-order strong coupling (full line)
and our data (dotted line).}
\end{center}
\vspace{-1.0cm}
\end{figure}
%%% Ende Figur %%%
%
In the low-energy domain $\gbsq$ is dominated by non-perturbative 
contributions.
Since the boundary fields are locally pure gauge configurations, any 
dependence of the effective action $\Gamma$ on the background field is 
caused by correlations around the spatially periodic lattice.
At large volume they are exponentially suppressed in a theory with a 
mass gap, and one expects
\be
\gbsq(L)\propto\exp\{m L\} \quad \mbox{at large $L$}\,.
\label{gb_largeL}
\ee
The mass $m$ is characteristic for the gauge field dynamics close to 
the boundaries (but not obviously related to the bulk correlation 
length).
Based on this argument we attempt to compare the behaviour of 
$\gbsq$ as a function of $L$ with the leading-order prediction from 
the strong-coupling expansion \cite{mG_StrCpl}:
$\gbsqi(L) \sim A\times\exp\{-m(\beta)L\}$, with 
$m=(3/4)\,m_{\rm G}=-3\ln\beta$ and $m_{\rm G}$ the $0^{++}$ 
glueball mass.
Thus, assuming this ansatz, we fit:
\be
\ln\left\{\gbsq(L)\right\}=
A'+\left(m\lmax\right)\times\left(L/\lmax\right)\,.
\ee
Fig.~\ref{StrFig} confronts the outcome of this analysis applied to 
the two highest points with the slope obtained in leading-order strong 
coupling,
\be
m r_0=\left(3/4\right)\times\left(m_{\rm G}r_0\right)
\,\,\Leftrightarrow\,\,
m\lmax \approx 2.3\,,
\ee
with $m_{\rm G}r_0\approx 4.3$ from \cite{mG_teper}.
Exploring the possibility, whether in eq.~(\ref{gb_largeL}) a prefactor
$L^c$, $c\neq0$, can build up when higher orders in the strong-coupling 
expansion are summed, would demand a calculation of at least a few 
further orders.
%
%%%%%%%%%%%%%%%%%%%%%%%%%%%%%%%%%%%%%%%%%%%%%%%%%%%%%%%%%%%%%%%%%%%%%%%%%
%
% bibliography
%

%

\begin{thebibliography}{99}
%
\bibitem{gbar_SU3}
M. L\"uscher, R. Sommer, P. Weisz and U. Wolff,
Nucl. Phys. B413 (1994) 481.
%
\bibitem{gbar_Nf2}
ALPHA Collaboration, A. Bode et al.,
Phys. Lett. B515 (2001) 49, hep-lat/0105003.
%
\bibitem{ZP} 
S. Capitani, M. L\"uscher, R. Sommer and H. Wittig, 
Nucl. Phys. B544 (1999) 669.
%
\bibitem{NPren_lat00}
S.~Sint,
Nucl.\,\,Phys.\,\,Proc.\,Suppl.\,94 (2001) 79.
%
\bibitem{RFSST}
M. L\"uscher, P. Weisz and U. Wolff,
Nucl. Phys. B359 (1991) 221.
%
\bibitem{APE_lat01}
APE Collaboration, these proceedings;
see also: \textsf{http://www-zeuthen.desy.de/ape}\,\,.
%
\bibitem{SF}
M. L\"uscher, R. Narayanan, P. Weisz and U. Wolff,
Nucl. Phys. B384 (1992) 168.
%
\bibitem{r0_ALPHA}
M. Guagnelli, R. Sommer and H. Wittig,
Nucl. Phys. B535 (1998) 389.
%
\bibitem{ct_2loop}
A. Bode, P. Weisz and U. Wolff, 
Nucl. Phys. B576 (2000) 517.
%
\bibitem{r0_contd}
S. Necco and R. Sommer, hep-lat/0108008. 
%
\bibitem{mG_StrCpl}
I. Montvay and G. M\"unster,
{\it Quantum Fields on a Lattice};
P. Weisz, priv. communication.
%
\bibitem{mG_teper}
M. J. Teper, hep-th/9812187.
%
\end{thebibliography}
\end{document}